\documentclass[final,english]{bullsrsl}[2022/06/15]

\usepackage[latin1]{inputenc}
\usepackage[T1]{fontenc}

\usepackage{natbib} 
\usepackage{graphicx}

\begin{document}
\title{Investigating the role of pre-supernova massive stars in the acceleration of galactic cosmic rays}

\author[affil={1},corresponding]{Micha\"el}{De Becker}
\author[affil={2}]{Santiago}{del Palacio}
\author[affil={3}]{Paula}{Benaglia}
\author[affil={4}]{Anandmayee}{Tej}
\author[affil={5}]{Benito}{Marcote}
\author[affil={3}]{Gustavo Esteban}{Romero}
\author[affil={6}]{Valenti}{Bosch-Ramon}
\author[affil={7}]{C.H.}{Ishwara-Chandra}
\affiliation[1]{Space sciences, Technologies and Astrophysics Research (STAR) Institute, University of Li\`ege, Belgium}
\affiliation[2]{Department of Space, Earth and Environment, Chalmers University of Technology, Gothenburg, Sweden}
\affiliation[3]{Instituto Argentino de Radioastronom\'ia, Villa Elisa, Buenos Aires, Argentina}
\affiliation[4]{Indian Institute for Space science and Technology (IIST), Thiruvananthapuram, India}
\affiliation[5]{Joint Institute for VLBI in Europe (JIVE), Dwingeloo, The Netherlands}
\affiliation[6]{Departament de F\'{i}sica Qu\`antica i Astrof\'{i}sica, Institut de Ci\`encies del Cosmos (ICC), Universitat de Barcelona, Spain}
\affiliation[7]{National Centre for Radio Astrophysics (NCRA), Pune, India}
\correspondance{Michael.DeBecker@uliege.be}
\date{30th April 2023}
\maketitle


%

\begin{abstract}
Galactic cosmic rays (GCRs) constitute a significant part of the energy budget of our Galaxy, and the study of their accelerators is of high importance in modern astrophysics. Their main sources are likely supernova remnants (SNRs). These objects are capable to convert a part of their mechanical energy into accelerated charged particles. However, even though the mechanical energy reservoir of SNRs is promising, a conversion rate into particle energy of 10 to 20\% is necessary to feed the population of GCRs. Such an efficiency is however not guaranteed. Complementary sources deserve thus to be investigated. This communication aims to address the question of the contribution to the acceleration of GCRs by pre-supernova massive stars in binary or higher multiplicity systems.
\end{abstract}


\section{Sources of Galactic cosmic rays}
Galactic Cosmic Rays constitute a significant component of our galactic environment. From the point of view of energy budget considerations, the GCR reservoir is close to equipartition with the magnetic field, the radiation field and the kinetic/turbulent energy of interstellar gas \citep[e.g.][]{Draine2011}. There is a consensus about the Galactic origin of cosmic rays at energies below a few PeV, i.e. below the so-called {\it knee} of the cosmic ray spectrum. The uniform slope in that region of the spectrum strongly suggests a consistent process is at the origin of their acceleration, and diffusive shock acceleration (DSA, \citealt[e.g.][]{Drury1983}) is the most likely candidate. Cosmic ray accelerators have thus to be identified among astrophysical environments capable to drive strong shock, with the availability of an abundant mechanical energy reservoir (large amount of material and high bulk velocities). In this context, the most obvious candidates are supernova remnants (see Section\,\ref{SNR}). They provide indeed a substantial energy reservoir where cosmic ray energy can be extracted, and their capability to accelerate particles up to quite high energies has been demonstrated notably by the detection TeV $\gamma$-rays confirming the acceleration of cosmic ray protons \citep{Ackermann2013}. It is, however, worth investigating potential complementary sources. The capability of SNRs to accelerate GCRs up to PeV energies is still perfectly clear, and recent results suggest that pulsar wind nebulae may provide a valuable contribution close to the {\it knee} \citep{LiuWang2021,Peng2022}. On the other hand, it is also worth considering the role of massive stars in their pre-supernova phase, i.e. across their full evolution before the supernova explosion, as they also constitute valuable energy reservoirs capable to accelerate particles. In particular, we will focus on massive stars in multiple systems that offer appropriate conditions for DSA to operate (see Section\,\ref{preSNR}). 

\subsection{Supernova remnants (SNRs)}\label{SNR}
Supernovae of any type inject a substantial amount of mechanical energy in the interstellar medium (ISM). These events are at the origin of supernova remnants: large amounts of material are expelled at high speed into the surrounding medium. These outflows are highly supersonic, and constitute prominent laboratories for shock physics in the ISM. Despite the diversity of precursors of SNRs (runaway fusion of a critical white dwarf, core collapse of massive stars of various evolutionary masses), all of them release a typical mechanical energy of the order of 10$^{51}$\,erg. Depending on the total mass of stellar material expelled by the supernova (typically of the order of a few M$\odot$) this leads to ejecta with velocities of a few 10$^3$ km\,s$^{-1}$, or up to about 10$^4$ km\,s$^{-1}$. The expanding shell consists mainly in a forward shock characterized by a high Mach number, capable to efficiently drive DSA. Assuming a supernova rate of 2.5 -- 3.0 events per century in the Milky Way, this translates into a total mechanical power of the order of 10$^{42}$ erg\,s$^{-1}$ \citep{Vink2020}. Assuming now a leaky box model with GCR energy loss dominated by escape, a steady state population would require a cosmic ray acceleration power of 1--2 $\times$ 10$^{41}$ erg\,s$^{-1}$. The requirement for SNRs to be the only contributor to GCR acceleration provided they can inject 10 to 20$\%$ of their mechanical energy into DSA \citep{GS1964}. Even though it sounds reasonable to assume that the acceleration efficiency should be high for young SNRs, i.e. when the shock velocity is still quite high, they contribute less at an advanced stage of the Sedov-Taylor phase and later when the shock velocity decreases. The uncertainties on the rate at which SNRs inject energy into DSA and the rather short duration of the high speed phase of its evolution open the possibility they may not be the only contributor. 

\subsection{Particle Accelerating Colliding-Wind Binaries (PACWBs)}\label{preSNR}

Binary or higher multiplicity systems made of hot, massive stars (O-type, early B-type, and their evolved counterparts such as Luminous Blue Variable and Wolf-Rayet stars) harbour typical precursors of core collapse supernovae. These stars are characterized by strong stellar winds with terminal velocities $V_\infty$ = 1000 -- 3000 km\,s$^{-1}$, and high mass loss rates (${\dot M}$ = 10$^{-7}$ -- 10$^{-5}$ M$_\odot$\,yr$^{-1}$). These quantities altogether express the wind kinetic power (i.e. mechanical energy conveyed per unit time, $\frac{1}{2}\,{\dot M}\,V_\infty^2$), that can reach values as high as 10$^{38}$ erg\,s$^{-1}$ in the most extreme cases. This kinetic power is made available for various physical processes, including non-thermal ones of importance for the topic of cosmic ray astrophysics. 

In multiple systems, stellar winds collide and produce high Mach number shocks, with pre-shock velocities typically equal to $V_\infty$. Such systems constitute thus adequate environments for shock physics. DSA is also likely responsible for the acceleration of particles up to relativistic velocities, as evidenced in about 50 systems, hence the Particle-Accelerating Colliding-Wind Binary status (PACWBs, \citealt{pacwbcata}, \url{www.astro.uliege.be/~debecker/pacwb}). Most particle accelerators among massive binaries are identified through the detection of synchrotron radio emission \citep[e.g.][]{Abbott1984,Benaglia2006,Dougherty2005,Blomme2007,Marcote2021}. The local magnetic field active in the synchrotron emission region (coincident with the colliding-wind region), results likely from the amplification of a field of stellar origin \citep{DeBecker2007,delPalacio2023}. Let's note that most massive stars are found in multiple systems, allowing this scenario to operate in a potentially high fraction of their population \citep[e.g.][in this volume]{DA2023}.

The order of magnitude of the total kinetic energy released across the evolution of an individual massive star can be obtained by integrating its kinetic power over its evolution time scale. Considering appropriate time scales as a function of stellar mass and evolutionary stage \citep{Woosley2002}, one obtains total kinetic energies in the range of 10$^{50}$ to $4\times10^{51}$ erg. These numbers are in fair agreament with the results published by \citep{Seo2018}. One can thus conclude that massive stars do release about the same order of magnitude of cumulative kinetic energy through their stellar winds across their pre-supernova phases as during the supernova explosion. Interestingly enough, let's emphasize that for the whole duration of their evolution, massive binaries offer shock conditions that are globally steady from the point of view of their pre-shock velocity, i.e. winds collide at their terminal velocities for millions of years. PACWBs are thus efficient at accelerating particles over their full evolution time scale of millions of years, rather than at most a few millennia for SNRs.

\begin{figure}
\centering
\includegraphics[width=12cm]{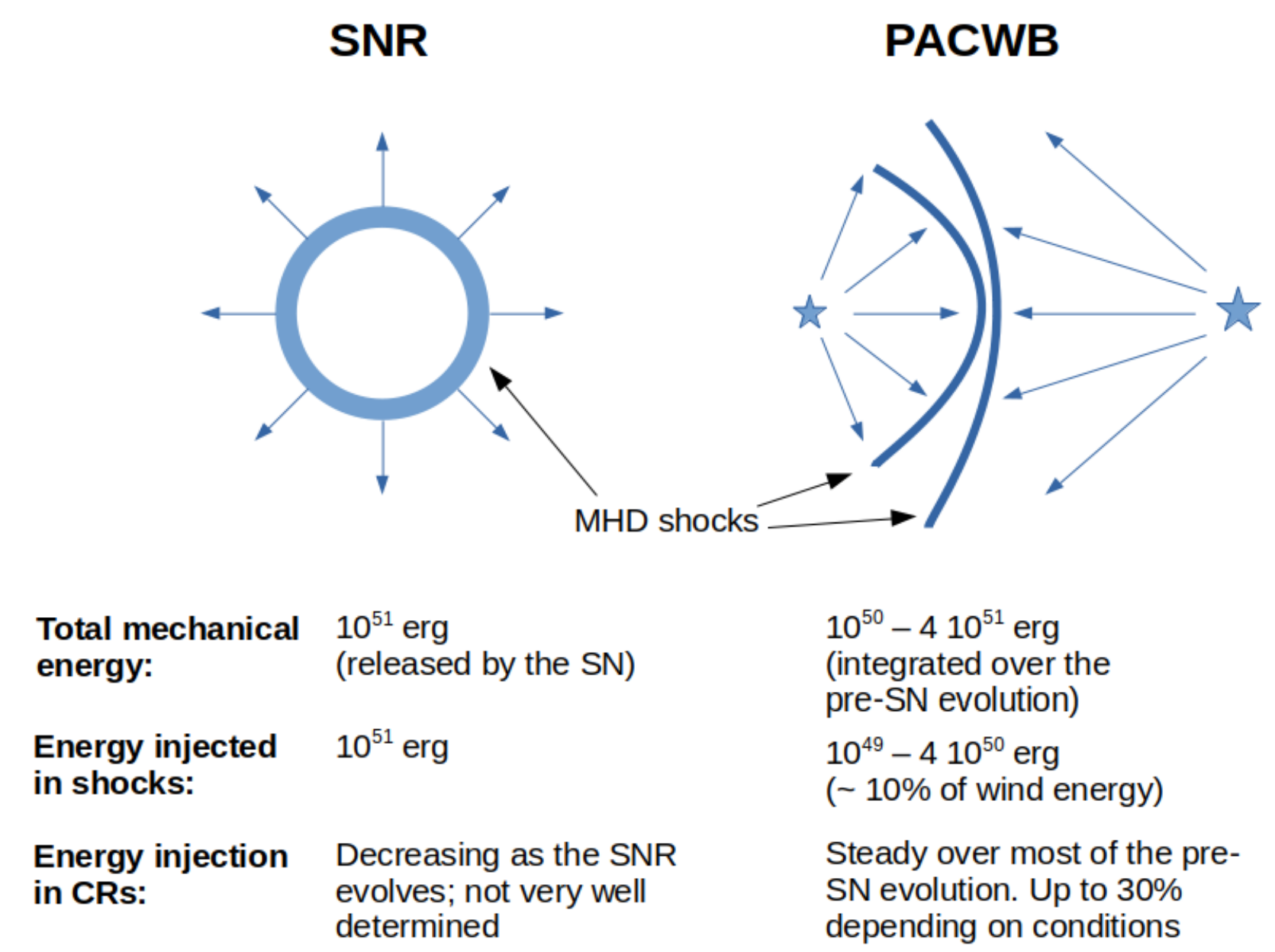}
\begin{minipage}{12cm}
\centering
\caption{Comparison of the participation of SNRs and PACWBs in the shock physics leading to the acceleration of GCRs. On the left, a spherical shell SNR is represented, assuming isotropic expansion for the sake of simplicity. On the right, the asymmetric illustration is meant to represent the usual case of non-equal stellar winds, where the colliding-wind region is pushed closer to the star producing the less powerful wind. Most known PACWBs are indeed made of stellar components of different spectral types and/or evolution stages, therefore characterized by different wind properties.}\label{Fig1}
\end{minipage}
\end{figure}

 If the wind kinetic energy has to be available to DSA, it needs to be  injected into the shocks in between the two stars. Only a fraction of the available energy will thus participate in shock physics, and consequently in DSA. Depending on the geometry of the system, let's assume that it amounts to about 10$\%$, even though it could be lower depending on the asymmetry of the system in term of kinetic power of the colliding stellar winds. It means that the effectively available mechanical energy could be in the range of 0.1--4 $\times$ 10$^{50}$ erg. From this point, the key question is that of the injection rate of energy into DSA. Recent models suggest that it may amount up to 30 $\%$ in some conditions \citep{Pittard2021}. This specific point requires a more intensive modelling of many individual systems, confronted to adequate measurements of their synchrotron radio emission. A summary of the comparison between SNRs and PACWBs from the point of view of their participation in GCR acceleration is provided in Fig.\,\ref{Fig1}.

\section{General strategy}
The identification of PACWBs can be made through the detection of synchrotron radiation. This requires the observations of a large number of targets to achieve a better view of the full population. The current census of PACWBs among massive binaries is severely underestimated because of strong observational biases \citep[e.g.][]{DeBecker2017}. Another requirement is to achieve a good census of binaries among massive stars. This is done notably through spectroscopic techniques, or high angular resolution imaging \citep{LeBouquin2017,Lanthermann2023}. Here again, the census of massive binaries is far from being complete. The combination of multiplicity studies of massive stars using various techniques, with intensive radio observations at several frequencies, is the key to address this question. This long term endeavour consists mainly in achieving a much better view of the population of particle-accelerating pre-supernova massive stars and characterize their main properties, with the objective to clarify their role as a contributor to the acceleration of GCRs. Even though synchrotron emission only tells us directly about the population of relativistic electrons, this is the best tracer of particle acceleration in massive star environments available to date. High energy observatories are still lacking the required sensitivity to address the non-thermal high-energy emission of a significant sample of massive stars at any stage before the supernova explosion. Let's however note a couple of notable exceptions such as Eta Car \citep[e.g.][]{Farnier2011} and WR\,11 \citep{MartiDevesa2020}, both identified as $\gamma$-ray sources. In addition, let's mention the case of Apep that turns out to be the only system with non-thermal emission detected both in the radio domain and in hard X-rays \citep{delPalacio2023}.

\section{Concluding remarks}

In this contribution, we provided a comparison between SNRs and pre-supernova PACWBs from the point of view of their likely contributions to the acceleration of GCRs. Stellar wind collisions in binary systems are known to accelerate particles, predominantly thanks to their synchrotron emission in the radio domain. Their stellar winds are energy reservoirs less abundant than SNRs, but they are capable to sustain a high acceleration efficiency for millions of years (to be compared to a few thousand years at most for SNRs). Many radio observations are needed to identify a significant number of particle accelerators among massive stars, and make a decisive leap forward in our understanding of their contribution to the production of GCRs. The relevant approach includes observations at various frequencies and various angular scales, to optimize the identification efficiency. The exploration of this scientific question is meant to achieve a more global view of the origin of GCRs, including both the pre- and post-supernova evolution of massive stars in our galaxy.

\begin{acknowledgments}
MDB would like to thank the organizers for the opportunity to present this poster, along with the Belgo-Indian Network for Astronomy and Astrophysics (BINA) for financial support. This work is part of the PANTERA-Stars initiative : \url{www.astro.uliege.be/~debecker/pantera}.
BM would like to thank the financial support from the State Agency for Research of the Spanish Ministry of Science and Innovation under grant PID2019-105510GB-C31/AEI/10.13039/501100011033 and through the ``Unit of Excellence Mar\'ia de Maeztu 2020-2023'' award to the Institute of Cosmos Sciences (CEX2019-000918-M). This research has made use of NASA's Astrophysics Data System Bibliographic Services. This work is supported by the Belgo-Indian Network for Astronomy and astrophysics (BINA), approved by the International Division, Department of Science and Technology (DST, Govt. of India; DST/INT/BELG/P-09/2017) and the Belgian Federal Science Policy Office (BELSPO, Govt. of Belgium; BL/33/IN12).
\end{acknowledgments}

\begin{furtherinformation}

\begin{orcids}
\orcid{0000-0002-1303-6534}{M.}{De Becker}
\orcid{0000-0002-5761-2417}{S.}{del Palacio}
\orcid{0000-0002-6683-3721}{P.}{Benaglia}
\orcid{0000-0001-9814-2354}{B.}{Marcote}

\end{orcids}

\begin{authorcontributions}
This work is part of a long term and collective effort where all co-authors provide contributions.
\end{authorcontributions}

\begin{conflictsofinterest}
The authors declare no conflict of interest.
\end{conflictsofinterest}

\end{furtherinformation}

\bibliographystyle{bullsrsl-en}

\bibliography{extra}

\end{document}